\documentclass[%
 article,twocolumn,prb,
 amsmath,amssymb,
 aps,
]{revtex4-1}
\usepackage[applemac]{inputenc}
\usepackage{mathtools}
\usepackage{slashed}
\usepackage{physics}
\usepackage{array}
\usepackage{amsmath}
\usepackage{color}
\usepackage{amsfonts}
\usepackage{amssymb}
\usepackage{graphicx}
\usepackage{amssymb,epsfig,subfigure}
\usepackage{hyperref}
\usepackage{comment}
\usepackage{tabularx}
\usepackage{bm}
\usepackage{dutchcal}
\usepackage{euscript}
\usepackage{graphicx}
\usepackage{color}
\usepackage [english]{babel}
\usepackage [autostyle, english = american]{csquotes}
\MakeOuterQuote{"}
\usepackage{amsfonts}
\usepackage{exscale}
\usepackage{amsbsy}
\usepackage{subfigure}
\usepackage{textcomp}
\usepackage{comment}
\usepackage{hyperref}
\usepackage{slashed}
\usepackage{tabularx}
\usepackage{euscript}
\usepackage{graphicx}
\usepackage{color}
\usepackage{amsfonts}
\usepackage{exscale}
\usepackage{amsbsy}
\usepackage{subfigure}
\usepackage{textcomp}
\usepackage{comment}
\usepackage{hyperref}
\usepackage{bm}
\usepackage{wrapfig}
\usepackage{mathrsfs}
\usepackage[font=footnotesize,labelfont=bf]{caption}

\def\ba#1\ea{\begin{align}#1\end{align}}
\def\bg#1\eg{\begin{gather}#1\end{gather}}
\def\bm#1\em{\begin{multline}#1\end{multline}}
\def\bmd#1\emd{\begin{multlined}#1\end{multlined}}

\newcommand{\beq}{\begin{eqnarray}}
\newcommand{\eeq}{\end{eqnarray}}
\begin{document}

\setcounter{page}1
\def\ppnumber{\vbox{\baselineskip14pt
}}

\def\ppdate{
} \date{\today}

\title{Corner Modes and Ground-State Degeneracy in Models with Gauge-Like Subsystem Symmetries}
\author{Julian May-Mann}
\affiliation{ \it Department of Physics and Institute for Condensed Matter Theory,\\  \it University of Illinois at Urbana-Champaign, \\  \it 1110 West Green Street, Urbana, Illinois 61801-3080, USA}
\author{Taylor L. Hughes}
\affiliation{ \it Department of Physics and Institute for Condensed Matter Theory,\\  \it University of Illinois at Urbana-Champaign, \\  \it 1110 West Green Street, Urbana, Illinois 61801-3080, USA}

\begin{abstract}

Subsystem symmetries are intermediate between global and gauge symmetries. One can treat these symmetries either like global symmetries that act on subregions of a system, or gauge symmetries that act on the regions transverse to the regions acted upon by the symmetry. We show that this latter interpretation can lead to an understanding of global, topology-dependent features in systems with subsystem symmetries. We demonstrate this with an exactly-solvable lattice model constructed from a 2D system of bosons coupled to a vector field with a 1D subsystem symmetry. The model is shown to host a robust ground state degeneracy that depends on the spatial topology of the underlying manifold, and localized zero energy modes on corners of the system. A continuum field theory description of these phenomena is derived in terms of an anisotropic, modified version of the Abelian K-matrix Chern-Simons field theory. We show that this continuum description can lead to geometric-type effects such as corner states and edge states whose character depends on the orientation of the edge.
\end{abstract}

\maketitle

\bigskip
\newpage



\section{Introduction}
\label{sec:1}
It is widely known that discrete gauge symmetries can give rise to topological order in 2+1D \cite{kogut1979,luscher1982,woit1983,kogut1983,wen1990,levin2003}. This began with work on 2+1D lattice gauge theory descriptions of quantum dimer models and resonating valence Bond states \cite{fradkin1979,fradkin1990,wen1991,moessner2001,ardonne2004,levin2006,fradkin2013}. Since then, there has been intense theoretical effort studying the properties of topological ordered lattice gauge theories \cite{jersak1983,brown1988,stack1994,engels1995,greensite2003,wang2011}. Key features of these systems include, a robust ground state degeneracy which depends on the topology of the underlying spatial manifold/lattice \cite{wen1989,misguich2002,levin2006}, fractionalized quasiparticles with unusual statistics  \cite{wen1991b,wen1991c,wen1991d,kitaev2006,nayak2008,levin2012}, and long-range entangled ground states \cite{kitaev2006b,chen2010,jiang2012}. 

A quintessential example of emergent topological order is Kitaev's toric code model, which realizes the deconfined phase of a $\mathbb{Z}_2$ lattice gauge theory \cite{kitaev1997,bravyi1998,kitaev2003,castelnovo2008,tupitsyn2010}. The model consists of a square lattice with spin-$1/2$ degrees of freedom defined on the links of lattice. The $\mathbb{Z}_2$ gauge transformation consists of flipping all spins around a single elementary plaquette. When defined on a manifold of genus $g$, the toric code system has a ground state degeneracy of $4^g$, which corresponds to the number of ways  $\mathbb{Z}_2$ gauge fluxes can be threaded through non-contractible loops in the system.

Recently, there has also been significant work in understanding the role of subsystem symmetries in topological phases of matter. For a $D$ dimensional system, subsystem symmetries (also refereed to as Gauge-Like symmetries) are sets of symmetries that act independently on $d$ dimensional subregions, with $0 <d<D$. Subsystem symmetries can be viewed as intermediate between gauge symmetries ($0$ dimensional subregions) and global symmetries ($D$ dimensional subregions).

In connection to topology, it has been shown that subsystem symmetries can lead to unique topological phases of matter known as subsystem symmetry protected topological  (SSPT) phases \cite{you2018}. SSPT phases have edge degrees of freedom that transform projectively under the subsystem symmetry. For open boundaries, SSPT's have a subextensive ground state degeneracy protected by the subsystem symmetries. In this way SSPT's are a subsystem generalization of (global) symmetry protected topological phases \cite{devakul2018}. 

Subsystem symmetries have also been studied in connection to fractonic phases of matter \cite{chamon2005,haah2011,nandkishore2018}. Fracton systems are 3+1D phases of matter, characterized by immobile excitations, or excitations which are confined to sub-dimensional regions. It has been found that \emph{gauging} a subsystem symmetry can lead to a fractonic phase\cite{bravyi2011,yoshida2013,williamson2016,vijay2016,you2018b,vijay2016}. Since fracton systems are believed to be described by rank 2 symmetric gauge theories, this field has also gained attention due to possible connections to elasticity and gravity theories \cite{pretko2017,pretko2018}.

Currently, the study of subsystem symmetries has been largely based on viewing a $d$-dimensional subsystem symmetry as a \emph{global} symmetry acting on $d$-dimensional subregions. However, there is also a complimentary view of a $d$-dimensional subsystem symmetry as a gauge symmetry acting on a $D-d$-dimensional subregion. For example, consider a $2d$ plane with coordinates $(x,y)$, where a subsystem symmetry acts along $1d$ $y = y_o$(const.) lines. Restricted to $y = y_o$ lines, the subsystem symmetry is a global symmetry. However, for $x = x_o$(const.) lines the subsystem symmetry is a local/gauge symmetry, since it only acts at the point $(x_o,y_o)$. 

Since subsystem symmetries behave like gauge symmetries in certain subregions, we believe that salient features of lattice gauge theories may occur in systems where the low energy physics is invariant under a subsystem symmetry. In particular we ask if subsystem symmetries can lead to interesting global phenomena in the same way that gauge symmetries do in topologically ordered phases. We answer this question in the affirmative by using a $D= 2$ model of bosons with a $d = 1$ $U(1)$ subsystem symmetry. Using two complimentary descriptions, we show that this model has multiple ground states on a torus, which cannot be locally distinguished. Furthermore, we show that for a rectangular system with open boundaries, there are gapless degrees of freedom that are localized to the system's corners.

This paper is organized as follows. In Section \ref{sec:2}, we construct the subsystem symmetry invariant model by using a coupled wire construction. In Section \ref{sec:3} we construct an effective projector Hamiltonian and use it to study the system. In Section \ref{sec:4} we construct and analyze a continuum description of the subsystem symmetry invariant model. In Section \ref{sec:5} we generalize the continuum description and discuss its features. Finally, we discuss and conclude these results in Section \ref{sec:6}. 

\section{Subsystem Symmetry Invariant Model}
\label{sec:2}
To construct our subsystem symmetry invariant model, we consider an array of complex bosonic wires on a square lattice with unit directions $\hat{x}$ and $\hat{y}$. The Hamiltonian for the wire array with wires aligned parallel to the $y$-direction is given by
\beq
H = -t\sum_{{\textbf{r}}} b^\dagger_{\textbf{r}+\hat{y}} b_{\textbf{r}} - \mu b^\dagger_{\textbf{r}} b_{\textbf{r}} + h.c.,
\label{eq:latmod1}
\eeq
where $b$ is a complex valued boson, and $\mu$ is a chemical potential. For a $L_x\times L_y$ lattice, this model has $L_x$ $U(1)$ symmetries which correspond to rotating the phase of a given wire. Formally, this symmetry operation is given by $b_{\textbf{r}}\rightarrow b_{\textbf{r}}e^{in\Lambda_{\textbf{r}}}$, where $n\in Z$, and $\Lambda_{\textbf{r}}$ is a real function that is constant along the $\hat{y}$ direction ($\Lambda_{\textbf{r}} = \Lambda_{\textbf{r}+\hat{y}}$). The factor of $n$ included in this definition is necessary for this system to have non-trivial features. 

We now want to couple these wires in such a way that the $L_x$ $U(1)$ subsystem symmetries are preserved. To do this, we will introduce a new set of fields $A$ defined on the links that connect sites $\textbf{r}$ and $\textbf{r}+\hat{x}$. These fields transform as $A_{\textbf{r},\textbf{r}+\hat{x}}\rightarrow A_{\textbf{r},\textbf{r}+\hat{x}} + (\Lambda_{\textbf{r}}- \Lambda_{\textbf{r}+\hat{x}})$ under the $U(1)$ subsystem symmetries. Introducing these fields, the Hamiltonian becomes
\beq
\nonumber H =  &-& t\sum_{{\textbf{r}}} b^\dagger_{\textbf{r}+\hat{y}}b_{\textbf{r}} -t' \sum_{{\textbf{r}} } b^\dagger_{\textbf{r}+\hat{x}} b_{\textbf{r}}e^{-in A_{\textbf{r},\textbf{r}+\hat{x}}}\\&-&\frac{K}{2} \sum_{\textbf{r}} e^{i(A_{{\textbf{r}+\hat{y},\textbf{r}+\hat{y}}+\hat{x}} - A_{{\textbf{r},\textbf{r}+\hat{x}}})}-\mu b^\dagger_{\textbf{r}} b_{\textbf{r}}+h.c.
\label{eq:latmod2}
\eeq
This model now has subsystem symmetries given by $b_{\textbf{r}}\rightarrow b_{\textbf{r}}e^{i n \Lambda_{\textbf{r}}}$ and $A_{\textbf{r},\textbf{r}+\hat{x}}\rightarrow A_{\textbf{r},\textbf{r}+\hat{x}} - ( \Lambda_{\textbf{r}} - \Lambda_{\textbf{r}+\hat{x}})$, where $\Lambda$ is a real function that is constant along the $\hat{y}$ direction. The $t'$ coupling in Eq. \ref{eq:latmod2} can be viewed as a subsystem generalization of a gauge connection, i.e., a way of coupling the bosons such that the subsystem symmetry is preserved. This coupling has also introduced vortex configurations where the value of $A$ jumps by $2\pi/n$. The term proportional to $K$ adds an energy cost to creating these vortices. Since the $K$ terms only couple fields that are neighbors in the $\hat{y}$-direction, these vortex excitations can only propagate along the $\hat{y}$-direction. 

To gain more insight into this Hamiltonian, let us restrict our attention to a line along the $\hat{x}$ direction defined as $l(y_o) = \{ \textbf{r} = (x,y)|y=y_o\}$ where $y_o$ is a constant. Let us extract the section of the Hamiltonian that acts \textit{only} on $l(y_o)$. The resulting $1d$ Hamiltonian for this subregion is 
\beq
H_{1d} = &\sum_{x}&\left( -t' b^\dagger_{x+1} b_{x} e^{-in A_{x,x+1}} - \mu b^\dagger_{x} b_{x} + h.c.\right),
\label{eq:gaugemod1}
\eeq
where $x\equiv (y_o,x)$. This is exactly the Hamiltonian for $1d$ charge $n$ bosons coupled to a gauge field $A$. The gauge transformations are given by $b_{x}\rightarrow b_{x}e^{in\Lambda'_{x}}$ and $A_{x,x+1}\rightarrow A_{x,x+1} + (  \Lambda'_{x}- \Lambda'_{x+1})$. This is exactly the subsystem transformation of the full system restricted to the $l(y_o)$ line.  So, along the $l(y_o)$ subregion, the subsystem symmetry corresponds to a $1d$ gauge symmetry.

Motivated by this, we can consider the expectation value of the Wilson loops of the dimensionally reduced $1d$ system $W_{1d} = \exp(i \sum_{x} A_{x,x+1})$. For periodic boundary conditions, the expectation value of $W_{1d}$ can be changed by a factor of $e^{i2\pi /n}$ by threading a unit of flux through the $1d$ system. In terms of the $A$ fields, the flux threading sends $A_{x,x+1} \rightarrow A_{x,x+1} + 2\pi /(nL_x)$, for each $x$. In the full $2d$ system, $W_{1d}$ becomes the operator $W_{l(y_o)} = \exp(i \sum_{\textbf{r}\in l(y_o)} A_{\textbf{r},\textbf{r}+\hat{x}})$. This operator is invariant under the $U(1)$ subsystem symmetries of Eq. \ref{eq:latmod1}. For periodic boundaries in the $\hat{x}$ direction, we can also define a "flux insertion" operation that sends $A_{\textbf{r},\textbf{r}+\hat{x}} \rightarrow A_{\textbf{r},\textbf{r}+\hat{x}} + 2\pi /(nL_x)$ for each ${\textbf{r}}$. This will change the expectation value of $W_{l(y_o)}$ by a factor of $e^{i2\pi /n}$.

It is clear that $W_{l(y_o)}$ is similar to the Wilson loops of a $2d$ lattice gauge theory. To illustrate the similarities and differences between lattice gauge theories and Eq. \ref{eq:latmod2}, let us consider these systems on a torus. For a $2d$ lattice gauge theory there are two distinct non-contactable Wilson loops: one oriented in the $\hat{x}$ direction, and one oriented in the $\hat{y}$ direction. The expectation value of these loops can be changed by threading flux through the $\hat{y}$ or $\hat{x}$ directions respectively. However, for Eq. \ref{eq:latmod2}, the Wilson loop-like operator $W_{l(y_o)}$ is fixed to be oriented in the $\hat{x}$ direction. As a result, the system only responds to threading flux through the $\hat{y}$ direction. Motivated by this, it will prove useful to think of Eq. \ref{eq:latmod2} as a gauge theory where the Wilson loops are restricted to be oriented in the $\hat{y}$ direction, or equivalently where flux can only be inserted in the $\hat{x}$ direction.

Now let us tune $\mu$ such that there is a large boson occupancy per site. $b$ can then be replaced with the rotor variable $e^{i\theta}$, where $\theta$ corresponds to the phase of the complex boson $b$\cite{phillips2012}. The Hamiltonian then becomes
\beq
\nonumber H =   &-&t\sum_{{\textbf{r}}} e^{i(\theta_{\textbf{r}} -\theta_{\textbf{r}+\hat{y}})}-t'\sum_{ {\textbf{r}}} e^{i(\theta_{\textbf{r}} -\theta_{\textbf{r}+\hat{x}} - n A_{{\textbf{r},\textbf{r}+\hat{x}}})}\\ &-&  \frac{K}{2} \sum_{\textbf{r}} e^{i(A_{{\textbf{r}+\hat{y},\textbf{r}+\hat{y}}+\hat{x}} - A_{{\textbf{r},\textbf{r}+\hat{x}}})} + h.c..
\label{eq:HiggsHam}
\eeq
The subsystem symmetry is now given by $A_{{\textbf{r},\textbf{r}+\hat{x}}} \rightarrow A_{{\textbf{r},\textbf{r}+\hat{x}}} + (\Lambda_{\textbf{r}} - \Lambda_{\textbf{r}+\hat{x}})$, and $\theta_{\textbf{r}} \rightarrow \theta_{\textbf{r}} + n\Lambda_{\textbf{r}}$ where $\Lambda_{\textbf{r}}$ is constant along the $\hat{y}$ direction. This model is the main result of this section. 

It is worth noting that due to the generalized Elitzur's theorem \cite{batista2005}, the continuous $1d$ subsystem symmetry of Eq. \ref{eq:HiggsHam} cannot be spontaneously broken. So the ground state of Eq. \ref{eq:HiggsHam} must be invariant under under all subsystem symmetry transformations, as must all local observables. This is similar to gauge theories, where the ground state and local observables must also be invariant under all local gauge transformations.

\section{Effective Projector Hamiltonian}
\label{sec:3}
To better study Eq. \ref{eq:HiggsHam}, it will be useful to construct an effective description in terms of an exactly solvable model of commuting projectors. The resulting model will be non-local, however it will be useful to determine key features of Eq. \ref{eq:HiggsHam} such as ground state degeneracy, and edge physics. In Section \ref{sec:4}, we will rederive these results using a local continuum description of Eq. \ref{eq:HiggsHam}. 

We will consider the case where $t,t' \rightarrow \infty$ while $K$ remains finite. The low energy excitations will thereby be violations of the term proportional to $K$ (vortices of $A$) in Eq. \ref{eq:HiggsHam}. To be explicit, let us consider an effective description for $n = 2$. The vortices of $A$ will therefore be $\pi$-vortices, where $\exp(i A) \rightarrow -\exp(i A)$. In the large $t'$ limit we can rewrite $A$ as
\beq
A_{\textbf{r},\textbf{r}+\hat{x}} = \frac{1}{2}(\theta_{\textbf{r}}-\theta_{\textbf{r}+\hat{x}})+\alpha_{\textbf{r},\textbf{r}+\hat{x}},
\label{eq:AIdent}
\eeq
where $\alpha_{\textbf{r},\textbf{r}+\hat{x}}$ is a $\pi$-valued variable ($\alpha$ only takes on values of $0$ or $\pi$) that corresponds to the vortices of the $A$ field. Let us now examine how these fields transform under a subsystem symmetry transformation given by $\Lambda$ satisfying $\Lambda_{\textbf{r}}=\Lambda_{\textbf{r}+\hat{y}}$. It will be useful to decompose $\Lambda \equiv \Lambda^s + \Lambda^\pi$, where  $\Lambda^s$ takes on values in $\lbrack 0,\pi)$ and $ \Lambda^\pi$ is a $\pi$-valued function. Under such a transformation 
\beq
\nonumber A_{\textbf{r},\textbf{r}+\hat{x}} &\rightarrow & A_{\textbf{r},\textbf{r}+\hat{x}} + (\Lambda^s_{\textbf{r}}-\Lambda^s_{\textbf{r}+\hat{x}})+(\Lambda^\pi_{\textbf{r}}-\Lambda^\pi_{\textbf{r}+\hat{x}})\\ \theta_{\textbf{r}}& \rightarrow & \theta_{\textbf{r}} + 2\Lambda^s_{\textbf{r}} + 2\Lambda^\pi_{\textbf{r}} = \theta_{\textbf{r}} + 2\Lambda^s_{\textbf{r}}
\label{eq:AIdent2}
\eeq
where we have used the fact that $\theta$ is $2\pi$ periodic. Comparing Eq. \ref{eq:AIdent} and \ref{eq:AIdent2}, we see that the transformation law for $\alpha$ is $\alpha_{\textbf{r},\textbf{r}+\hat{x}} \rightarrow \alpha_{\textbf{r},\textbf{r}+\hat{x}} +(\Lambda^\pi_{\textbf{r}}-\Lambda^\pi_{\textbf{r}+\hat{x}})$. So $\alpha$ is only acted on by transformation generated by $\Lambda^\pi$. Since $2\Lambda^\pi = 0\mod(2\pi)$, the transformations generated by $\Lambda^\pi$ form a $\mathbb{Z}_2$ subgroup of the full $U(1)$ group of subsystem symmetry transformations. 

\begin{figure}
\centering
\includegraphics[width=\linewidth*3/5]{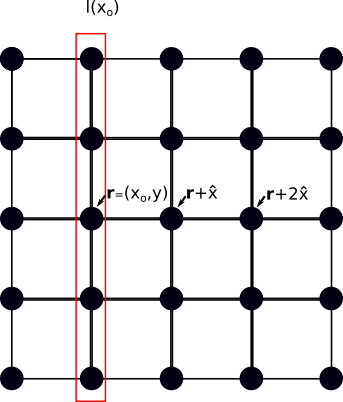}
\caption{A column of sites $l(x_o)$ (red).}
\label{fig:GaussDia}
\end{figure}

Because $\alpha$ is $\pi$-valued, we can identify $\exp(i \alpha) = \sigma^z$, where $\sigma^z$ is a Pauli matrix. Using Eq. \ref{eq:AIdent}, the Hamiltonian Eq. \ref{eq:HiggsHam} becomes 
\beq
\nonumber H =  &-&  \frac{K}{2}\sum_{ {\textbf{r}}} \sigma^z_{\textbf{r}+\hat{y},\textbf{r}+\hat{y}+\hat{x}}\sigma^z_{\textbf{r},\textbf{r}+\hat{x}}e^{\frac{1}{2}(\theta_{\textbf{r}}-\theta_{\textbf{r}+\hat{x}} - \theta_{\textbf{r}+\hat{y}}+\theta_{\textbf{r}+\hat{y}+\hat{x}})}\\ &-& t\sum_{ {\textbf{r}}} e^{i(\theta_{\textbf{r}} -\theta_{\textbf{r}+\hat{y}})}+h.c. .
\label{eq:HiggsHamEff1}
\eeq The aforementioned $\mathbb{Z}_2$ subsystem symmetry generated by $\Lambda^\pi$ flips the spins $\sigma^z_{\textbf{r},\textbf{r}+\hat{x}} \rightarrow -\sigma^z_{\textbf{r},\textbf{r}+\hat{x}}$ on an even number of columns. In terms of the spin variables, this symmetry transformation is generated by $G[l(x_o)] = \prod_{\textbf{r}\in l(x_o)}  \sigma^x_{\textbf{r},\textbf{r}+\hat{x}} \sigma^x_{{\textbf{r}+\hat{x},\textbf{r}+2\hat{x}}}$, where $l(x_o) = \{ \textbf{r} = (x,y)|x=x_o\}$ (see Fig. \ref{fig:GaussDia}).

The full Hilbert space of Eq. \ref{eq:HiggsHamEff1} is spanned by $\bigotimes_{\textbf{r}}\ket{\bar\sigma^z_{\textbf{r},\textbf{r}+\hat{x}}}\ket{\bar\theta_{\textbf{r}}}$. These are eigenstates with eigenvalues $\sigma^z_{\textbf{r},\textbf{r}+\hat{x}}\ket{\bar\sigma^z_{\textbf{r},\textbf{r}+\hat{x}}} = \bar\sigma^z_{\textbf{r},\textbf{r}+\hat{x}}\ket{\bar\sigma^z_{\textbf{r},\textbf{r}+\hat{x}}}$ ($\bar\sigma^z \in \pm 1$) and $ \theta_\textbf{r}\ket{\bar\theta_{\textbf{r}}} = \bar\theta_{\textbf{r}}\ket{\bar\theta_{\textbf{r}}}$ ($\bar{\theta}\in \lbrack 0,2\pi)$). In the $t\rightarrow \infty$ limit, we will only consider states that satisfy $\bar{\theta}_{\textbf{r}} = \bar{\theta}_{
\textbf{r}+\hat{y}}$. Using this the Hamiltonian becomes 
\beq
H = -K\sum_{{\textbf{r}} }  \sigma^z_{{\textbf{r} +\hat{y} ,\textbf{r}}+\hat{y}+\hat{x}} \sigma^z_{{\textbf{r},\textbf{r}+\hat{x}}}.
\label{eq:HiggsHam2}
\eeq In this limit the phase fluctuations are frozen out energetically and the effective model acts on the \textit{restricted} Hilbert space spanned only by the spin operators $\sigma^z$. Formally this is a mapping that takes a state $\bigotimes_{\textbf{r}}\ket{\bar\sigma^z_{\textbf{r},\textbf{r}+\hat{x}}}\ket{\bar\theta_{\textbf{r}}}\rightarrow \bigotimes_{\textbf{r}}\ket{\bar\sigma^z_{\textbf{r},\textbf{r}+\hat{x}}}.$

Additionally, due to the generalized Elitzur's theorem, all observables must be invariant under the $U(1)$ subsystem symmetries. Because of this, we should focus on just the "physical subspace" of this reduced Hilbert space, which consists of states that are invariant under the $U(1)$ subsystem symmetries generated by $\Lambda$. Under the aforementioned mapping, the physical subspace of the \emph{full} Hilbert space maps to a subspace of the \emph{restricted} Hilbert space that is invariant under the $\mathbb{Z}_2$ subsystem symmetry subgroup that acts on $\exp(i\alpha) = \sigma^z$. To project the restricted Hilbert space onto the corresponding physical subspace, we note that a subsystem symmetry invariant state $\ket{\psi}$ will satisfy $G[l(x_o)]\ket{\psi} = \ket{\psi}$ for all columns $l(x_o)$. This condition can be enforced in the low-energy subspace by adding the term $-JG[l(x_o)]$ (with $J>0$) to the Hamiltonian Eq. \ref{eq:HiggsHam2}. The resulting effective projector Hamiltonian is 
\beq
H_{eff} = &-& K\sum_{{\textbf{r}} }  \sigma^z_{{\textbf{r} +\hat{y} ,\textbf{r}}+\hat{y}+\hat{x}} \sigma^z_{{\textbf{r},\textbf{r}+\hat{x}}}\\ &-& J\sum_{x_o} \prod_{\textbf{r}\in l(x_o)}\sigma^x_{{\textbf{r},\textbf{r}+\hat{x}}} \sigma^x_{{\textbf{r}+\hat{x},\textbf{r}+2\hat{x}}}.
\eeq 
The low energy sector will now be invariant under the subsystem symmetry. The second term in this Hamiltonian is notably non-local. This is an artifact of projecting to the physical Hilbert space. Nevertheless, this effective model provides a simple and useful description that we can use to study the low energy features of the full system Eq. \ref{eq:HiggsHam}.

\begin{figure}
\centering
\includegraphics[width=\linewidth/2]{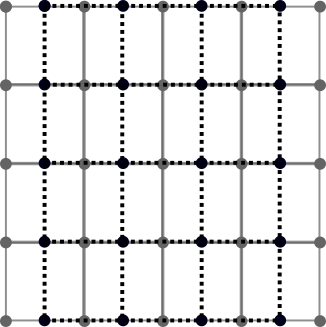}
\caption{The original Lattice (grey), where the $A$ fields are defined on $\hat{x}$ oriented links, and the new lattice (black) where the $A$ fields are defined on sites.}
\label{fig:LatticeFig1}
\end{figure}

 It will now be useful to simplify the lattice on which we have defined this effective spin model. Let us define a new lattice such that the sites of the new lattice are the links connecting the sites $\textbf{r}$ and $\textbf{r}+{\hat{x}}$ of the original lattice. This means that the $A$ fields now live on sites instead of links. The new lattice is shown Fig. \ref{fig:LatticeFig1}. After switching to the new lattice the Hamiltonian simplifies to 
\beq
H = - K \sum_{\textbf{r}} \sigma^z_{\textbf{r}} \sigma^z_{{\textbf{r}}+\hat{y}} - J\sum_{x_o}\prod_{\textbf{r} \in l(x_o)}  \sigma^x_{\textbf{r}} \sigma^x_{{\textbf{r}}+\hat{x}}.
\label{eq:2HOHam}
\eeq 
where ${\textbf{r}}$ are the sites on the new lattice, and $\hat{x}$ and $\hat{y}$ are now the unit directions of the new lattice. $l(x_o) = \{\textbf{r} = (x,y)| x = x_o \}$ is now the set of spins along a given straight line in the $\hat{y}$ direction. 

\begin{figure}
\centering
\includegraphics[width=\linewidth*4/5]{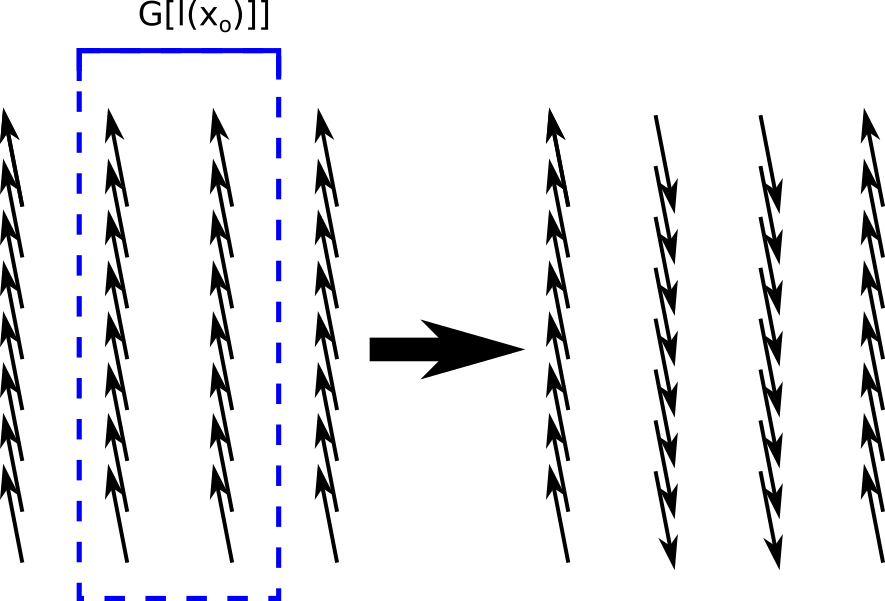}
\caption{The action of the $\mathbb{Z}_2$ subsystem symmetry, $G[l(x_o)]$ which flips all spins on a pair of neighboring columns.}
\label{fig:SymDia}
\end{figure}

This spin model is the main result of this section. All terms in the Hamiltonian commute, and so the spin model is exactly solvable. The subsystem symmetry here is generated by 
\beq
G[l(x_o)] = \prod_{\textbf{r} \in l(x_o)}  \sigma^x_{\textbf{r}} \sigma^x_{{\textbf{r}}+\hat{x}}.
\label{eq:SSGen1}
\eeq
This operation is shown in Fig. \ref{fig:SymDia}.  As we can see, the non-local second term in Eq. \ref{eq:2HOHam} guarantees that the ground state of the system is invariant under this transformation. Eq. \ref{eq:2HOHam} also has a second subsystem symmetry generated by \beq
G[l(y_o)] = \prod_{\textbf{r} \in l(y_o)}  \sigma^z_{\textbf{r}} \sigma^z_{{\textbf{r}}+\hat{y}},
\label{eq:SSGen2}
\eeq
where $l(y_o) = \{\textbf{r} = (x,y)| y=y_o \}$ is a line of spins in the $\hat{x}$ direction. Due to the first term in Eq. \ref{eq:2HOHam}, the ground state will be invariant under this second subsystem symmetry as well. 

Eq. \ref{eq:2HOHam} is similar to the quantum compass model\cite{kugel1973}, a precursor to the Kitaev honeycomb model\cite{kitaev2003}, which is given by the Hamiltonian 
\beq
H_{compass} = - J_z \sum_{\textbf{r}} \sigma^z_{\textbf{r}} \sigma^z_{{\textbf{r}}+\hat{y}} - J_x\sum_{\textbf{r}} \sigma^x_{\textbf{r}} \sigma^x_{{\textbf{r}}+\hat{x}}.
\label{eq:QCM}
\eeq 
Indeed, the quantum compass model and the spin model Eq. \ref{eq:2HOHam} share the same subsystem symmetries, and Eq. \ref{eq:2HOHam} can also arise as the effective description of the $J_z > J_x$ phase of Eq. \ref{eq:QCM} in finite sized systems. In this case, the effective $K$ will be proportional to $ (J_x/J_z)^{L_y}$. However, despite the apparent similarities, these models have different ground state properties in the thermodynamic limit. It is known that the quantum compass model has 2 phases corresponding to $J_x>J_z$ and $J_z>J_x$\cite{dorier2005}. In both phases, the number of ground states scales as $2^L$ for an $L\times L$ system. The $J_x = J_z$ point marks a first order phase transition that connects these two phases\cite{chen2007}. In contrast, the spin model Eq. \ref{eq:2HOHam} has a gapped phase with a finite number of ground states, even in the thermodynamic limit. This will be shown in the following sections.

\subsection{Ground States and Excitations}

The ground state of the effective spin model Eq. \ref{eq:2HOHam} can be found by minimizing each of the commuting terms. We can intuitively understand the nature of the ground state in the following way. The terms proportional to $K$ in Eq. \ref{eq:2HOHam} describe an array of decoupled Ising chains. Thus, for $J = 0$, the spin model is simply an array of Ising chains in the ferromagnetic phase. In the low-energy subspace, each chain can then be characterized by a single magnetization variable $\bar{\sigma}^z_{x_o} = \langle  \sigma^z_{\textbf{r}}\rangle_{\textbf{r} \in l(x_o)}$. 

The terms proportional to $J$ in Eq. Eq. \ref{eq:2HOHam} flip all spins on a pair of the neighboring Ising chains (see Fig. \ref{fig:SymDia}), i.e., each term flips a pair of magnetizations, e.g.,  $\bar{\sigma}^z_{x_o}$ and $\bar{\sigma}^z_{x_o+\hat{x}}$. Let us define the operator $\bar{\sigma}^x_{x_o} = \prod_{\textbf{r}\in l(x_o)}\sigma^x$. Since $\bar{\sigma}^z_{x_o} = \langle  \sigma^z_{\textbf{r}}\rangle_{\textbf{r} \in l(x_o)}$, $\bar{\sigma}^x_{x_o}\bar{\sigma}^z_{x_o} = -\bar{\sigma}^x_{x_o}\bar{\sigma}^z_{x_o}$. In terms of $\bar{\sigma}$, the Hamiltonian Eq. \ref{eq:2HOHam} becomes
\beq
H = - J\sum_{x_o} \bar{\sigma}^x_{x_o}\bar{\sigma}^x_{x_o+\hat{x}}.
\label{eq:Heff}
\eeq
This Hamiltonian is just another ferromagnetic Ising chain, with the ferromagnetism oriented in the $x$-direction. So the effect of the term proportional to $J$ in Eq. \ref{eq:2HOHam} is to orient the magnetization of the original Ising chains. In particular, if we start with a ground state for $J = 0$, we can determine the ground state for $J > 0$ by acting on the $J=0$ ground state with the operator 
\beq
D_s = \prod_{l(x_o)}\frac{1}{2}\left(1+\prod_{\textbf{r} \in l(x_o)}  \sigma^x_{\textbf{r}} \sigma^x_{{\textbf{r}}+x}\right).
\eeq 
To see this, consider a state $\ket{\psi}$ that minimizes Eq. \ref{eq:2HOHam} with $J=0$. Then  $\sigma^z_{\textbf{r}} \sigma^z_{{\textbf{r}}+y}\ket{\psi} = \ket{\psi}$ for all $\textbf{r}$. Since $\sigma^z_{\textbf{r}} \sigma^z_{{\textbf{r}}+y}D_s\ket{\psi} = D_s\sigma^z_{\textbf{r}} \sigma^z_{{\textbf{r}}+y}\ket{\psi} = D_s\ket{\psi}$, $D_s\ket{\psi}$ minimizes all $K$ terms in Eq. \ref{eq:2HOHam}. It is also true that $(\prod_{\textbf{r} \in l(x_o)}  \sigma^x_{\textbf{r}} \sigma^x_{{\textbf{r}}+\hat{x}})D_s = D_s$, for all $x_o,$ and by extension, $(\prod_{\textbf{r} \in l(x_o)}  \sigma^x_{\textbf{r}} \sigma^x_{{\textbf{r}}+\hat{x}})D_s \ket{\psi}= D_s\ket{\psi}$. So $D_s\ket{\psi}$ also minimizes all $J$ terms in Eq. \ref{eq:2HOHam}. $D_s\ket{\psi}$ thereby minimizes the entire Hamiltonian with $J>0,$ and is the ground state. 

We note here that $D_s$ is in fact exactly the projection operator that projects the restricted Hilbert space of Eq. \ref{eq:HiggsHam2} to the subsystem symmetry invariant physical subspace of the restricted Hilbert space. As we shall demonstrate below, the number of ground states will depend on the topology of the lattice. The excited states of the spin model are characterized by having either $\sigma^z_{\textbf{r}} \sigma^z_{{\textbf{r}}+\hat{y}} = -1$ or $\prod_{\textbf{r} \in l(x_o)}  \sigma^x_{\textbf{r}}\sigma^x_{\textbf{r}+\hat{x}} = -1$, which have an excitation energy of $2K$ and $2J$ respectively.   

\subsection{Ground State Degeneracy}

\begin{figure}
\includegraphics[width=\linewidth/2]{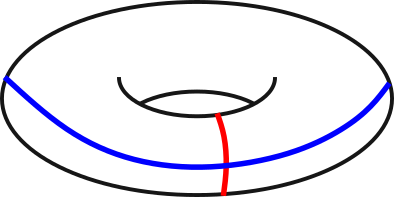}
\caption{ (Loops operators which commute with the Hamiltonian Eq. \ref{eq:2HOHam} on a torus. Red lines are $\sigma^z$ strings and blue lines are $\sigma^x$ strings}
\label{fig:LoopsFig}
\end{figure}

A key feature of the subsystem symmetry invariant model Eq. \ref{eq:HiggsHam} is the existence of multiple ground states that cannot be locally distinguished. We will demonstrate this by considering the effective spin model Eq. \ref{eq:2HOHam} on a torus. To find the number of ground states, we will identify operators that commute with the Hamiltonian, and use them to label the degenerate ground states. The non-trivial operators that commute with the Hamiltonian Eq. \ref{eq:2HOHam} are
\beq
\nonumber W_{l(y_o)} = \prod_{{\textbf{r}} \in l(y_o)}\sigma^z_{\textbf{r}}\\ W_{l(x_o)} = \prod_{{\textbf{r}} \in l(x_o)}\sigma^x_{\textbf{r}},
\label{eq:strings}
\eeq
where $l(y_o) = \{\textbf{r} = (x,y)|y = y_o \}$ is a closed loop in the $\hat{x}$ direction, and $l(x_o) = \{\textbf{r} = (x,y)|x = x_o \}$ is a closed loop in the $\hat{y}$ direction. On a torus, the $l(y_o)$ and $l(x_o)$ loops will be the two cycles that define the torus. These loops are shown in Fig. \ref{fig:LoopsFig}. For an $L_x \times L_y$ torus, the total number of $W_{l(y_o)}$ operators is $L_y$ and the number of $W_{l(x_o)}$ operators is $L_x$. Since $W^2_{l(x_o)} = W^2_{l(y_o)}=1$, both loop operators are $\mathbb{Z}_2$ operators. We can identify the symmetry operator $G[l(x_o)]$ as the product of the neighboring loop operators $W_{l(x_o)}$ and $W_{l(x_o+\hat{x})}$, and similarly identify $G[l(y_o)]$ as the product of $W_{l(y_o)}$ and $W_{l(y_o+\hat{y})}$. 

All loops $l(x_o)$ and $l(y_o)$ on the torus intersect once, and so \textit{all} $W_{l(y_o)}$ operators anti-commute with \textit{all} $W_{l(x_o)}$ operators. The minimum dimension needed to represent this anti-commuting algebra is 2, leading to 2 distinct ground states. If we were to diagonalize the ground state subspace to label them by their $W_{l(y_o)}$ eigenvalue, then the 2 ground states would be related by the acting on a given ground state with an operator $W_{l(x_o)}$. Since $W_{l(y_o)}$ and $W_{l(x_o)}$ are both non-local operators, this degeneracy is robust to local perturbations. 

The degeneracy can also be found by counting the number of constraints for $L_x\times L_y$ spins on a torus. Let us first consider the terms proportional to $K$ in Eq. \ref{eq:2HOHam}. These terms describe a system of $L_x$ Ising chains with periodic boundaries. Each chain contributes $L_y-1$ unique constraints, leading to $L_x(L_y-1)$ unique constraints from the $K$ terms in Eq. \ref{eq:2HOHam}. The terms proportional to $J$ in Eq. \ref{eq:2HOHam} then give $L_x-1$ unique constraints. Since all terms in Eq. \ref{eq:2HOHam} commute, all these constraints can be simultaneously satisfied, leading to $L_x(L_y-1) + L_x -1 = L_x\times L_y - 1$ constraints in total. There is thereby $1$ net free spin degree of freedom which corresponds to the $2$ ground states that were previously identified. 

It is useful to compare these results to the case of a $\mathbb{Z}_2$ lattice gauge theory on a torus. In $\mathbb{Z}_2$ lattice gauge theory models, there are 2 additional ground states on a torus (for a total of 4 ground states)\cite{kitaev2006}. These 2 additional ground states occur since non-contractible loops of $\sigma^x$ operators oriented in the $\hat{x}$ direction, and non-contractible loops of $\sigma^z$ oriented in the $\hat{y}$ direction also commute with the $\mathbb{Z}_2$ lattice gauge theory Hamiltonian, and anti-commute with each other. These operators do not commute with the spin model Eq. \ref{eq:2HOHam}, and so the number of ground states is reduced to 2. 

On a sphere all string operators $W_{l(y_o)}$ and $W_{l(x_o)}$ commute, and so the ground state of Eq. \ref{eq:2HOHam} on a sphere is unique. We also show this explicitly in Appendix A by counting constraints. This topology-dependent degeneracy is reminiscent of the topological ground state degeneracy found in topological ordered systems, though it is important to note that our spin model has a non-local constraint. The non-locality will be removed when we discuss the continuum limit.

\subsection{Open Boundaries and Corner Modes}
We shall now consider the system with open boundaries. For simplicity, we shall take the lattice to be an $L_x \times L_y$ rectangle with open boundaries. For this geometry, the terms proportional to $K$ in Eq. \ref{eq:2HOHam} give $L_x(L_y-1)$ constraints, and the terms proportional to $J$ give $L_x-1$ constraints, leading to $L_x\times L_y -1$ constraints which can be simultaneously satisfied. There is then a single free spin 1/2 degree of freedom, leading to 2 ground states. 

In the string picture, this can be seen by the anti-commutation between the zero energy operators $W_{l(y_o)}$ and $W_{l(x_o)}$ from Eq. \ref{eq:strings} where $l(y_o)$ (resp. $l(x_o)$) is now a string in the $\hat{x}$ (resp. $\hat{y}$) direction stretching from one boundary to the other. Since the system has open boundaries, the string operators do not have to form closed loops to commute with the Hamiltonian and be invariant under the subsystem symmetries of the model. Since all $W_{l(y_o)}$ operators anti-commute with all $W_{l(x_o)}$ operators there are degenerate 2 ground states.

Furthermore, these 2 ground states correspond to anti-commuting corner degrees of freedom. To show this, we will switch to a Majorana representation of the spin-1/2 degrees of freedom\cite{kitaev2003}. This is done by introducing 4 Majorana degrees of freedom at each site $\gamma^1$, $\gamma^2$, $\gamma^3$, and $\gamma^4$. The spin degrees of freedom then become $\sigma^x = i\gamma^1\gamma^2$, $\sigma^y= i\gamma^1\gamma^3$, and $\sigma^z = i\gamma^1\gamma^4$, with the local constraint that $\gamma^1\gamma^2\gamma^3\gamma^4 = 1$. Setting $K = J$, the spin model can be described by the mean field Hamiltonian
\beq
H_{MF} = -\sum_{\langle \textbf{r}\textbf{r'}\rangle}\sum_{i,j}J_{ij,\textbf{r}\textbf{r'}}\gamma^i_{\textbf{r}}\gamma^j_{\textbf{r'}},
\label{eq:MajMF}
\eeq
where $J_{ij,\textbf{r}\textbf{r'}}$ corresponds to the Majorana dimerization pattern given in Fig \ref{fig:4MajFig}. The construction of this mean field Hamiltonian is outlined in Appendix B.  The ground state of the spin model can then be found by projecting the mean field ground state of this Majorana Hamiltonian onto the physical states using the projector $P = \sum_{\textbf{r}}\frac{1}{2}(1+\gamma^1_{\textbf{r}}\gamma^2_{\textbf{r}}\gamma^3_{\textbf{r}}\gamma^4_{\textbf{r}})$, and then using the above identification between the spin operators and Majorana fermions.

\begin{figure}
\includegraphics[width=\linewidth/2]{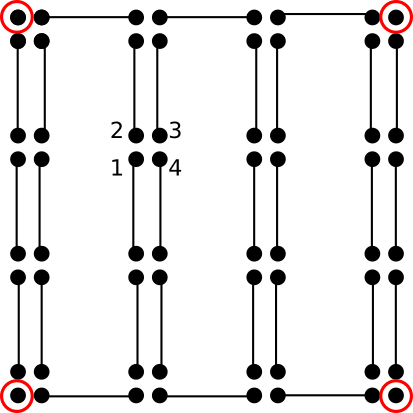}
\caption{ The mean field majorana representation of the spin model Eq. \ref{eq:2HOHam}. Black lines indicate a non-zero $J_{ij,\textbf{r}\textbf{r'}}$ in Eq. \ref{eq:MajMF} and dimerized pair of Majorana fermions. Zero energy Majorana corner modes are circled in red.}
\label{fig:4MajFig}
\end{figure}

For the rectangular geometry, there are 4 free Majorana degrees of freedom located at the corners of the lattice (see Fig. \ref{fig:4MajFig}). This leads to 4 ground states, but only 2 are physical after projecting onto the physical states. The reduction in ground states can be viewed as a consequence of the full model being bosonic, i.e., fermion parity even. For the rectangular geometry, the degrees of freedom for the spin model are thereby zero energy corner operators leading to a robust ground state degeneracy. 
We note that this particular dimerization pattern and corner mode configuration is similar to an insulator model presented in Ref. \onlinecite{benalcazarPRB} that has corner charge, but vanishing quadrupole moment. Indeed, the corner modes considered here are similar to what is found in higher order topological insulators\cite{benalcazar2017,langbehn2017,schindler2018,khalaf2018}. However, for the spin model Eq. \ref{eq:2HOHam}, the corner operators are non-local. This is because individual Majorana corner operators do not commute with the projector operator $P$. Only pairs of Majorana corner operators commute with $P$ and are physical.

\section{Continuum Theory}
\label{sec:4}
We now seek a complementary continuum description of Eq. \ref{eq:HiggsHam}. First, we note that Eq. \ref{eq:HiggsHam} is the low energy description of 
\beq
\nonumber H = &-&t\sum_{\langle {\textbf{r}\textbf{r'}} \rangle } \cos(\theta_{\textbf{r}} -\theta_{\textbf{r'}} - n A_{{\textbf{r}\textbf{r'}}}) - K \sum_p \cos(F_p)\\ &-& \bar{m}^2 \sum_{ \textbf{r}} \cos(A_{\textbf{r},\textbf{r}+\hat{y}}),
\label{eq:latmod3}
\eeq
where $A$ fields are now defined along both $\hat{x}$ and $\hat{y}$ oriented links, and $\langle \textbf{r}\textbf{r'} \rangle$ are neighboring sites. The sum over $p$ is over plaquettes with corners $i,j,k,l,$ and $F_p = A_{i,j}-{A_{k,l}}+A_{i,k}-A_{j,l}$. In the low energy ($\bar{m}^2 \rightarrow \infty$) limit, $A_{\textbf{r},\textbf{r}+\hat{y}}$ is pinned to be $0$ by the cosine term. Upon substituting this into Eq. \ref{eq:latmod3}, the Hamiltonian reduces to Eq. \ref{eq:HiggsHam} with $t' = t$. This model has the subsystem symmetry given by $A_{\textbf{r},\textbf{r'}} \rightarrow A_{\textbf{r},\textbf{r'}} + (\Lambda_{\textbf{r}} - \Lambda_{\textbf{r'}})$, and $\theta_{\textbf{r}} \rightarrow \theta_{\textbf{r}} + n\Lambda_{\textbf{r}}$ where $\Lambda_{\textbf{r}} = \Lambda_{\textbf{r}+\hat{y}}$. This is the same symmetry as in Eq. \ref{eq:HiggsHam}. We note that $A_{\textbf{r},\textbf{r}+\hat{y}}$ is \emph{invariant} under these transformations. Eq. \ref{eq:latmod3} is the Hamiltonian for bosons minimally coupled to a vector field $A$ with an additional mass term for the fields $A$ oriented along the $y$ direction. It is worth explicitly stating that this model is \textit{not} a gauge theory due to the additional mass term for $A_{\textbf{r},\textbf{r}+\hat{y}}$.

 The continuum description of Eq. \ref{eq:latmod3} in Euclidean space is
\beq
\nonumber\mathcal{L}_{E} = &\frac{1}{4}&F^{\mu \nu }F_{\mu \nu } + \rho (\partial_t \theta - n A_t)^2 + \rho(\partial_x \theta - n A_x)^2\\  &+& \rho(\partial_y \theta - n A_y)^2 + \bar{m}^2 A^2_y- A_\mu j^\mu,
\label{eq:ContHiggs1}
\eeq
where $\rho$ is a constant, and we have included a current $j^\mu$ that couples to the fluctuations of the $A$ field. To study the dynamics of the phase $\theta$, we will introduce the variable $a_\mu = -\frac{1}{n}\partial_\mu \theta$, and shift $A_{\mu} \rightarrow A_{\mu} - a_{\mu}$. After this Eq. \ref{eq:ContHiggs1} becomes 
\beq
\nonumber \mathcal{L}_E = &\frac{1}{4}&F^{\mu \nu }F_{\mu \nu } + \rho n^2 (A_t)^2 + \rho n^2 (A_x)^2+M^2 a_y^2 \\  &-&  2\bar{m}^2 A_ya_y  + \bar{m}^2 A_y^2 - A_\mu j^\mu - a_\mu j^\mu,
\eeq
where we have introduced $M^2 \equiv \bar{m}^2 + \rho n^2$. Since the $A$ field is now massive, it can be integrated out, leaving a theory just in terms of $a$. The $a$ field is now also coupled to $j$, causing the excitations of the $a$ field to have a corresponding current $j$. As desired, this model has a subsystem symmetry where $a_{\mu} \rightarrow a_{\mu} + \partial_{\mu} \Lambda_a(x,t)$ and $\Lambda_a$ is function of $x$ and $t$ \textit{only}.  

From the equations of motion for $a_y$, we have that $a_y \propto \frac{1}{M^2}$. As a result, in the low energy limit where $M^2 \rightarrow \infty$, $a_y$ vanishes. In this limit, $j^y$ (the current in the $y$ direction) is removed from the theory, and the only current is in the $x$ direction ($j^x$). This means that the excitations of the $a$ field will only move in the $x$ direction. This is a form of sub-dimensional dynamics, where the excitations are only able to move in certain lower dimensional subregions. 

We will also need to consider the vortex dynamics of the $a$ field. This is done by introducing a vortex current $\tilde{j}^\mu$, and setting $\tilde{j}^\mu = \frac{in}{2\pi} \epsilon^{\mu \nu \lambda}\partial_\nu a_\lambda$. To enforce this constraint, we will introduce the field $b_\mu$ as a Lagrange multiplier for Eq. \ref{eq:ContHiggs1}
\beq
\mathcal{L}_E \rightarrow \mathcal{L}_E+ b_\mu \tilde{j}^\mu -  \frac{in}{2\pi} b_\mu \epsilon^{\mu \nu \lambda}\partial_\nu a_\lambda. 
\label{eq:LagM1}
\eeq
In this construction, there are vortex currents in both the $x$ and $y$ directions ($\tilde{j}^x$ and $\tilde{j}^y$). However, in the original lattice model Eq. \ref{eq:HiggsHam}, the vortices were only able to move in the $y$ direction. To remedy this, we will add the term $M^2b^2_x$. The equation of motion for $b_x$ then gives that $b_x\propto \frac{1}{M^2}$, and in the low energy limit ($M^2 \rightarrow \infty$) $b_x \rightarrow 0$. In this limit, the $\tilde{j}^x$ vortex current is removed from the theory, and there is only a vortex current in the $y$-direction ($\tilde{j}^y$). As a result, the vortices of the $a$ field are confined to move in the $y$-direction as in the lattice model.

After adding the $b$ field, integrating out the massive $A$ field, and keeping only the long wavelength contributions, the Lagrangian density becomes
\beq
\nonumber \mathcal{L}_E = &\frac{in}{2\pi}& b_\mu \epsilon^{\mu \nu \lambda}\partial_\nu a_\lambda + M^2 a_y^2 +M^2 b^2_x\\ &-& b_\mu \tilde{j}^\mu - a_\mu j^\mu,
\label{eq:ContHiggs2}
\eeq
This model is the main result of this section. It is worth noting that this theory has the form of a mutual Chern-Simons theory with additional mass terms for $a_y$ and $b_x$ \cite{bardeen1965,hansson2004,diamantini2006}. This observation will be allow us to generalize this model in Section \ref{sec:5}. 

In Eq. \ref{eq:ContHiggs2}, it is also apparent that there is a second subsystem symmetry where $b_{\mu} \rightarrow b_{\mu} + \partial_{\mu} \Lambda_b(y,t)$ and $\Lambda_b$ is only a function of $y$ and $t$. This is the same as the subsystem symmetry generated by $G[l(y_o)]$ (Eq. \ref{eq:SSGen2}) in the effective projector Hamiltonian.

\subsection{Ground State Degeneracy}
We will now calculate the ground state degeneracy of the continuum model on a torus. To do this, we will first rotate back to Minkowski space, and set the currents $j=\tilde{j} = 0$,
\beq
\mathcal{L} = \frac{n}{2\pi} b_\mu \epsilon^{\mu \nu \lambda}\partial_\nu a_\lambda + M^2a_y^2 + M^2b_x^2. 
\label{eq:LagLE}
\eeq
From the equations of motion for $a_y$ and $b_x$, we have that $a_y \propto b_x \propto \frac{1}{M^2}$. At low energies, $a_y \rightarrow 0$ and $b_x \rightarrow 0$, and the action becomes
\beq
\mathcal{S} = \frac{n}{2\pi}  \int d^3 x  (b_y  \partial_t a_x - a_t \partial_x b_y - b_t \partial_y a_x).
\label{eq:ContHiggs3}
\eeq
If we minimize the action with respect to $b_t$ and $a_t$ we find the equations of motion $\partial_y a_x = 0$ and $\partial_x b_y = 0$. On a torus, these equations of motion are solved by 
\beq
\nonumber a_x &=& \partial_x \phi(x,t) + \bar{a}_x(t)/L_x\\
b_y &=& \partial_y \theta(y,t) + \bar{b}_y(t)/L_y.
\eeq
Here $\phi$ is a function of $x$ and $t$ only and is periodic on the torus, $\theta$ is a function of $y$ and $t$ and is periodic on the torus, and $ \bar{a}_x$ and $\bar{b}_y$ are functions of $t$ only. $L_{x,y}$ are the length dimensions of the torus. 

After substituting these terms into Eq. \ref{eq:ContHiggs3} and integrating over the $x$ and $y$ coordinates, the action reduces to
\beq
\mathcal{S} = \frac{n}{2\pi}  \int dt ( \bar{b}_y  \partial_t  \bar{a}_x).
\eeq
Using canonical commutation relations, we have that $[\bar{b}_y,\bar{a}_x] = i 2\pi/n$. Since $\bar{b}_y$ and $\bar{a}_x$ are $2\pi$ periodic variables, the observables are $\exp(i \bar{b}_y)$ and $\exp(i \bar{a}_x)$, which obey the commutation relationship, 
\beq
e^{i \bar{b}_y} e^{i \bar{a}_x} = e^{i\frac{2\pi}{n}}e^{i \bar{a}_x}e^{i \bar{b}_y}.
\eeq
In order to satisfy this operator algebra, there must be $n$ ground states. This is consistent with what was found using the effective projector model with $n = 2$. We note that for a conventional mutual Chern-Simons theory the ground state degeneracy would be $n^2.$
 
\subsection{Corner Modes}
To find the edge degrees of freedom for a system with open boundaries we will use the low energy description with no external currents in Minkowski space (Eq. \ref{eq:ContHiggs3}). For a rectangular system with open boundaries, the equations of motion for $a_x$ and $b_y$ are solved by 
\beq
\nonumber a_x = \partial_x\phi(x,t)\\
b_y = \partial_y \theta(y,t). 
\eeq
Using this, the action becomes
\beq
\nonumber \mathcal{S} &=& \int d^3 x \frac{n}{2\pi}\partial_y \theta(y,t) \partial_t \partial_x\phi(x,t)\\
&=& \int d^3 x \frac{n}{2\pi}\partial_y \partial_x \lbrack\theta(y,t) \partial_t \phi(x,t)\rbrack,
\eeq
which is a total derivative for both $x$ and $y$. If the system is defined on the rectangle $x_0 \leq x \leq x_1$ and $y_0 \leq y \leq y_1$, the action becomes 
\beq
\nonumber \mathcal{S} &=& \frac{n}{2\pi}\int dt (\theta(y_1,t) - \theta(y_0,t)) \partial_t (\phi(x_1,t) - \phi(x_0,t)).\\
\label{eq:CornerAct}
\eeq
This action describes localized operators $\eta_{j,k} \equiv \exp(i \phi(x_j,t) - i\theta(y_k,t))$, which are defined at the corners of the system $(x_j,y_k)$. Since the Hamiltonian corresponding to the action Eq. \ref{eq:CornerAct} vanishes, the $\eta_{i,j}$ are zero energy operators. It should be noted that there is a redundancy in the corner mode description, since $\eta_{0,0}\eta_{1,1}\eta^\dagger_{1,0}\eta^\dagger_{0,1} = 1$. 

Using the canonical commutation relationships from Eq. \ref{eq:CornerAct}, the $\eta_{j,k}$ operators satisfy the algebra 
\beq
\eta_{j,k}\eta_{j',k'}=\eta_{j',k'}\eta_{j,k}\exp\left(\frac{2\pi i [j'\cdot k-j\cdot k']}{n} \right).
\eeq  Naively this would lead to $n^2$ ground states. However, if the constraint $\eta_{0,0}\eta_{1,1}\eta^\dagger_{1,0}\eta^\dagger_{0,1} = 1$ is taken in account there are actually only $n$ ground states. This agrees with what was found in using the effective model for $n=2$. As opposed to $2+1D$ Abelian Chern-Simons field theories, where the edge theory is a $1+1D$ CFT \cite{balachandran1991,dunne1999,wen2004,fujita2009,lu2012}, the edge theory of the subsystem symmetry invariant model is given by $0+1D$ corner modes.

\section{Generalized Continuum Theory}
\label{sec:5}
To generalize the continuum description to include more vector fields, we note that Eq. \ref{eq:ContHiggs2} has the form of a Chern-Simons field theory with $K$-matrix $2\sigma^x,$ and mass terms for $a_y$ and $b_x$. Using this observation, we can generalize the continuum description of the subsystem symmetry invariant system by using the Lagrangian
\beq
\mathcal{L} = \frac{1}{4\pi} K^{IJ} \epsilon^{\mu \nu \lambda} a^I_{\mu} \partial_\nu a^J_{\lambda} + a^{I}_x M_x^{IJ} a^{J}_x + a^{I}_y M_y^{IJ} a^{J}_y,
\label{eq:ContGen}
\eeq
where $K$ is a $D\times D$ symmetric integer valued matrix. We will take $M_{x,y}$ to be diagonal with all entries either $m$ or $0$. As we shall see, in order for the canonical quantization to be consistent, we shall require the number of zero eigenvalues of $M_x$ and $M_y$ to be equal, i.e., dim(ker($M_x$)) = dim(ker($M_x$)) $= \mathcal{k} \leq D$. 

Minimizing the action with respect to $a^I_x$ and $a^I_y$ gives the equations of motion
\beq
M_x^{IJ} a^{J}_x = - \frac{1}{4\pi} K^{IJ} F_{ty}^J\\
M_y^{IJ} a^{J}_y= - \frac{1}{4\pi} K^{IJ} F_{xt}^J,
\eeq
where $F_{\mu \nu}^J = \epsilon^{\mu \nu}\partial_\mu a_{\nu}^J$. Let us write $M_{x,y}^{IJ} = m \times \bar{M}_{x,y}^{IJ}$, where $\bar{M}_{x,y}^{IJ} $ is a diagonal matrix of $1$'s and $0$'s. Using $\bar{M}$, the equations of motion are
\beq
\bar{M}_x^{IJ} a^{J}_x= - \frac{1}{4\pi m} K^{IJ} F_{ty}^J\\
\bar{M}_y^{IJ} a^{J}_y = - \frac{1}{4\pi m} K^{IJ} F_{xt}^J.
\eeq
At low energies $m \rightarrow \infty$, and 
\beq
\bar{M}_x^{IJ} a^{J}_x = 
\bar{M}_y^{IJ} a^{J}_y = 0.
\eeq
So all vector fields $a^{J}_{x,y}$ not in the respective kernels of $M_{x,y}$ are set to $0$ as $m \rightarrow \infty$. Setting $m\rightarrow \infty$ is thereby equivalent to projecting $a^{J}_{x,y}$ on to the respective $\mathcal{k}$ dimensional kernels of $M_{x,y}$. Since $M_{x,y}$ is a diagonal matrix, the kernel is spanned by a set of $\mathcal{k}$ = dim(ker($M_{x,y}$)) unit vectors. This means we can project onto the kernels of $M_{x,y}$ with $\mathcal{k}\times D$ matrices $V_{x,y}$, the rows of which are the unit vectors that span the kernels of $M_{x,y}$. 

The theory with $m\rightarrow \infty$ can then be expressed as follows. Define the reduced $K$ matrices as
\beq
\nonumber K_{tx}^{ij} &=& K_{xt}^{ji} \equiv K^{ik}V_x^{kj} \\
\nonumber K_{ty}^{ij} &=& K_{yt}^{ji} \equiv K^{ik}V_y^{kj}\\
K_{yx}^{ij} &=& K_{xy}^{ji} \equiv V_y^{il}K^{lk}V_x^{kj},
\eeq
and the vectors 
\beq
\nonumber \tilde{a}_x^i  &\equiv& V_x^{ij}a_x^j\\
\nonumber \tilde{a}_y^i  &\equiv& V_y^{ij}a_y^j\\
\tilde{a}_t^i  &\equiv& a_t^i.
\eeq
The effective Lagrangian in the $m\rightarrow \infty$ limit is then
\beq
\mathcal{L}_{eff} = \sum_{i,j}\sum_{ \mu,\nu,\lambda}\frac{1}{4\pi} \epsilon^{\mu \nu \lambda} K_{\mu \nu}^{ij} \tilde{a}_\mu^i \partial_\nu \tilde{a}_\lambda^j,
\label{eq:ContGen2}
\eeq
where we have explicitly included the sum for clarity. 

\subsection{Quantization}
In order to consistently, canonically quantize Eq. \ref{eq:ContGen2}, we need the following equations to be consistent 
\beq
\nonumber \sum_j K_{xy}^{ij} \lbrack \tilde{a}_x^k, \tilde{a}_y^j \rbrack &=& i2 \pi \delta^{i,k}\\
\sum_j K_{yx}^{ij} \lbrack \tilde{a}_y^k, \tilde{a}_x^j \rbrack &=& -i2\pi \delta^{i,k}.
\label{eq:QuantC1}
\eeq
To simplify this, we will define the matrix $A^{kj} = \lbrack \tilde{a}_x^k, \tilde{a}_y^j \rbrack$. Eq. \ref{eq:QuantC1} then becomes (using $K_{yx}^{ij} = K_{xy}^{ji}$)
\beq
\nonumber \sum_j K_{xy}^{ij} A^{kj} &=& i2 \pi \delta^{i,k}\\
\sum_j (K_{xy}^{T})^{ij} (A^{T})^{kj} &=&  i2\pi \delta^{i,k}.
\label{eq:QuantC2}
\eeq 

Let us consider the case where dim(ker($M_x$)) = $\mathcal{k}$ and  dim(ker($M_y$)) = $\mathcal{k}'$. Then $K_{xy} = K^T_{yx}$ is a $\mathcal{k} \times \mathcal{k}'$ matrix, and $A^{kj}$ is a $\mathcal{k}' \times \mathcal{k}$ matrix. Summing over the $i$ and $k$ indices in Eq. \ref{eq:QuantC2} gives 
\beq
\nonumber Tr(K_{xy}A) &=& i 2 \pi \mathcal{k} \\
Tr(K_{xy}^{T} A^{T}) &=& Tr(A K_{xy} ) =  i 2\pi \mathcal{k}'.
\label{eq:QuantC3}
\eeq
Since $Tr(K_{xy}A) = Tr(A K_{xy} )$, $\mathcal{k} = \mathcal{k}'$ in order for the quantization conditions to be consistent. This confirms our earlier assertion that we must have dim(ker($M_x$)) = dim(ker($M_y$)). 

If $\det(K_{xy})\neq 0$, Eq. \ref{eq:QuantC1} is solved by $\lbrack \tilde{a}_x^i, \tilde{a}_y^j \rbrack = 2\pi i (K_{xy}^{-1})^{ij}$. If $\det(K_{xy}) = 0$, the inverse of $K_{xy}$ will not be well defined, and the commutation relations will be ambiguous. Because of this, we shall assume that $\det(K_{xy})\neq 0$ from here on. It is worth noting that $K_{xy}$ must be square, but $K_{ty}$ and $ K_{tx}$ do not need to be square.

\subsection{Ground State Degeneracy}
We will now show that the ground state degeneracy on a torus is $|\det(K_{xy})|$ (which is valid because $K_{xy}$ is a square matrix). Let us minimize the action by setting the functional derivative of Eq. \ref{eq:ContGen2} with respect to $\tilde{a}_t^i$ equal to $0$. The resulting equations of motion are
\beq
K^{ij}_{ty} \partial_x \tilde{a}^j_y - K^{ij}_{tx} \partial_y \tilde{a}^j_x = 0.
\eeq
Since we are only concerned with global features of the system, we will use the solutions $\tilde{a}^i_{x,y} = \bar{a}^i_{x,y}/L_{x,y}$ where $\bar{a}^i_{x,y}$ is only a function of $t$, and $L_{x,y}$ are the lengths of the torus. Other solutions represent local fluctuations and do not contribute to global features.

Plugging these solutions into Eq. \ref{eq:ContGen2} and integrating over the $x$ and $y$ coordinates, we arrive at the action
\beq
\mathcal{S}= \int dt\frac{1}{2\pi}   K_{yx}^{ij} \bar{a}_y^i \partial_t \bar{a}_x^j .
\eeq From this we can calculate the algebra satisfied by the observables $\exp(i \bar{a}^i_{x,y})$. Using that $\lbrack \bar{a}^i_{x,y}, \bar{a}^j_{x,y} \rbrack = 2\pi i (K_{xy}^{-1})^{ij}$, the minimum dimension needed to satisfy the algebra of the $\exp(i \bar{a}^i_{x,y})$ operators is $|\det(K_{xy})|$. This leads to a ground state degeneracy of $|\det(K_{xy})|$ on a torus.

\subsection{Edge and Corner States and Example}
We will illustrate some of the ground state degeneracy and edge state possibilities using an example case. Consider a $4\times 4$ $K$-matrix, and mass matrices $M_x, M_y$ having kernel dimension equal to $2$. 
For the first example we will choose the $K$-matrix and mass matrices
\beq
\nonumber K &=& \begin{bmatrix} 0&2&0&1\\2&0&0&0\\0&0&4&0\\1&0&0&-4  \end{bmatrix}\\
\nonumber M_x &=& \begin{bmatrix} 0&0&0&0\\0&0&0&0\\0&0&m&0\\0&0&0&m  \end{bmatrix}\\
M_y &=& \begin{bmatrix} 0&0&0&0\\0&m&0&0\\0&0&m&0\\0&0&0&0  \end{bmatrix}.
\label{eq:MatEx}
\eeq
The corresponding fields will be labeled as $a^i_\mu$ with $i = 1,2,3,4$ and $\mu = (t,x,y)$. The $2\times 4$ $V_{x,y}$ matrices for the theory are 
\beq
\nonumber V_x &=& \begin{bmatrix} 1&0&0&0\\0&1&0&0 \end{bmatrix}\\
V_y &=& \begin{bmatrix} 1&0&0&0\\0&0&0&1 \end{bmatrix}.
\eeq

The reduced $K$ matrices are given by
\beq
K_{tx} &=& \begin{bmatrix} 0&2\\2&0\\0&0\\1&0 \end{bmatrix}\\\nonumber
K_{ty} &=& \begin{bmatrix} 0&1\\2&0\\0&0\\1&-4 \end{bmatrix}\\\nonumber
K_{yx} &=& \begin{bmatrix} 0&2\\1&0 \end{bmatrix}.
\eeq In the low energy limit, the Lagrangian density is given by 
\beq
\nonumber \mathcal{L} = \frac{1}{2\pi}[ &2& a^1_y\partial_t a^2_x + a^4_y\partial_t a^1_x - 2 a^1_t \partial_y a^2_x  - 2 a^2_t \partial_y  a^1_x\\\nonumber &-& a^4_t \partial_y a^1_x + a^1_t \partial_x a^4_y + 2a^2_t \partial_x a^1_y + a^4_t \partial_x a^1_y\\ &-& 4 a^4_t \partial_x a^4_y].
\label{eq:LagEx}
\eeq The canonical quantization commutation relations are $[a^1_y,a^2_x] = i\pi$ and
$[a^4_y,a^1_x] = i2\pi$. Minimizing the action with respect to $a^i_t$, we determine the equations of motion
\beq
\nonumber \partial_y a^1_x- \partial_x a^1_y&=& 0\\\nonumber \partial_y a^2_x &=& 0\\
\partial_x a^4_y &=& 0.
\label{eq:EQMGen}
\eeq

Let us consider the case where the system is put on a torus with side lengths $L_x$ and $L_y$. If we ignore local fluctuations of the fields, the equations of motion can be solved by $a^1_x = \bar{a}^1_x(t)/L_x$, $a^1_y = \bar{a}^1_y(t)/L_y$, $a^2_x = \bar{a}^2_x(t)/L_x$, and $a^4_y = \bar{a}^4_y(t)/L_y$. Since $\bar{a}^i_\mu$ is a $2\pi$ periodic variable, we will consider the operators $e^{i\bar{a}^i_\mu}.$ Using the canonical commutation relations, $ e^{i\bar{a}^y_x}$ and $e^{i\bar{a}^2_x}$ anti-commute, while all other terms commute. This means that there will be 2 ground states. This agrees with $|\det(K_{xy})| = 2$. 

Let us now consider the edge and corner modes of this system. The equations of motion Eq. \ref{eq:EQMGen} are solved by $a^1_{x,y} = \partial_{x,y}\phi^1(x,y,t)$, $a^2_x = \partial_x \phi^2(x,t)$, and $a^4_y = \partial_y \phi^4(y,t)$. Plugging these solutions into Eq. \ref{eq:LagEx}, the action becomes
\beq
\nonumber \mathcal{S} = \int d^3 x \frac{1}{2\pi} [&\partial_x& ( \partial_y \phi^4(y,t)  \partial_t\phi^1(x,y,t))\\&-& 2 \partial_y(  \partial_t \phi^1(x,y,t)\partial_x \phi^2(x,t))].
\eeq To illustrate the edge modes of this system, it will be useful to consider a half plane $x\leq 0$. If we assume that the fields vanish at spatial infinity, we can rewrite the action as
\beq
\mathcal{S} &=& \int dtdy \frac{1}{2\pi}\partial_t \phi^4(y,t) \partial_y \phi^1(0,y,t).
\label{eq:edge1}
\eeq  
This describes a non-chiral boson propagating along the $x=0$ edge. If we instead consider the $y\leq 0$ half plane the action is 
\beq
\mathcal{S} = & -&\int dtdx \frac{1}{\pi}   \partial_t \phi^1(x,0,t)\partial_x \phi^2(x,t).
\label{eq:edge2}
\eeq
This describes a \textit{different} non-chiral boson that propagates along the $y=0$ edge. To see that this is in fact a different non-chiral boson, we note that Eq. \ref{eq:edge1} and \ref{eq:edge2} describe a $U(1)_1$ and $U(1)_2$ non-chiral boson CFT respectively. 

It will also be useful to consider a quarter plane geometry $x\leq 0$ and $y\leq 0$. In this geometry, the action becomes
\beq
\nonumber \mathcal{S} &=& \int dtdy \frac{1}{2\pi}\partial_t \phi^4(y,t) \partial_y \phi^1(0,y,t)\\ & -&\int dtdx \frac{1}{\pi}   \partial_t \phi^1(x,0,t)\partial_x \phi^2(x,t).
\label{eq:edgeGen}
\eeq
As expected, the first line of Eq. \ref{eq:edgeGen} describes a non-chiral boson propagating along the $x=0, y\leq 0$ boundary, while the second line describes a different type of non-chiral boson propagating along the $x\leq0,y=0$ boundary. Since $\phi^1(0,y,t)$ and $\phi^1(x,0,t)$ coincide at the point $(x,y) = (0,0)$, the two non-chiral bosons are coupled at this point. This means that there will be a partially-transmitive  boundary connecting the two edges.  

Let us now consider a different example using the same $K$ matrix as in Eq. \ref{eq:MatEx}, but with the mass matrices 
\beq
\nonumber M_x &=& \begin{bmatrix} m&0&0&0\\0&0&0&0\\0&0&0&0\\0&0&0&m  \end{bmatrix}\\
M_y &=& \begin{bmatrix} 0&0&0&0\\0&m&0&0\\0&0&0&0\\0&0&0&m  \end{bmatrix}.
\eeq
Following the same analysis as before, \beq
K_{yx} = \begin{bmatrix}2&0\\0&4\end{bmatrix}.\eeq
This gives a ground state degeneracy of $|$det$(K_{yx})| = 8$. 

The equations of motion for these mass matrices are
\beq
\nonumber \partial_x a^1_y  &=& 0\\
\nonumber \partial_y a^2_x &=& 0\\
\partial_x a^3_y - \partial_y a^3_x &=& 0.
\eeq
The equations of motion are solved by $a^1_y = \partial_y \phi^1(y,t)$, $a^2_x = \partial_x \phi^2(x,t)$, and $a^3_{x,y} = \partial_{x,y} \phi^3(x,y,t)$. 

For a half plane with $x\leq 0$, the edge action becomes 
\beq
\nonumber \mathcal{S} = \int dtdy \frac{2}{\pi} \partial_t \phi^3(0,y,t) \partial_y \phi^3(0,y,t).
\eeq
This edge describes a $U(1)_4$ chiral boson. For a half plane with $y\leq 0$, the edge action is similarly 
\beq
\nonumber \mathcal{S} = \int dtdx \frac{2}{\pi} \partial_t \phi^3(x,0,t) \partial_x \phi^3(x,0,t),
\eeq
which describes the same $U(1)_4$ chiral boson. Finally, for the quarter plane $x\leq 0$, $y \leq 0$, the action is 
\beq
\nonumber \mathcal{S} &=& \int dtdy \frac{2}{\pi} \partial_t \phi^3(0,y,t) \partial_y \phi^3(0,y,t)\\\nonumber
&+&\int dtdx \frac{2}{\pi} \partial_t \phi^3(x,0,t) \partial_x \phi^3(x,0,t)\\
&+&\int dt \frac{1}{\pi} \phi^1(0,t) \partial_t \phi^2(0,t).
\label{eq:edge3}
\eeq
The first two parts of the edge action Eq. \ref{eq:edge3} describe the previously shown chiral edge modes. The final term describes a zero energy excitation $\exp(i [\phi^1(0,t)-\phi^2(0,t)])$ which is bound to the $(x,y) = (0,0)$ corner of the system. The boundary of this system thereby has coexisting propagating chiral edge modes, and localized corner modes. 

For a general theory given by Eq. \ref{eq:ContGen}, there will be corner modes if there exists $i$ and $j$ ($i \neq j$) such that $K^{ij} \neq 0$, $M^{ii}_x = m$, $M^{jj}_y = m$ $M^{jj}_x = 0$, and $M^{ii}_y = 0$. Setting $a^i_{y} = \partial_y \phi^i(y,t)$, and $a^j_{x} = \partial_x \phi^j(x,t)$, the localized mode for a corner at ($x_o,y_o$) is given by $\exp(i [\phi^i(y_o,t)-\phi^j(x_o,t)])$.

 These examples highlight the unusual edge/boundary physics that can occur in models of the form Eq. \ref{eq:ContGen}. In general, the edge theory for a given $K$ matrix and pair of mass matrices will consists of both propagating edge modes, and localized corner modes. Furthermore, for a given system, the edge theory may depend on the orientation of the edge since the theory can naturally support anisotropy.

\section{Discussion and Conclusion: Gauging Subsystem Symmetry and Open Problems}
\label{sec:6}
In this work we have shown that invariance under a subsystem symmetry can lead to a topology-dependent ground state degeneracy, and corner modes. We established this using both a exactly-solvable, but non-local, spin model, and a continuum field theory description. From this, we have shown that global, topology-dependent features can exist beyond the established paradigm of gauge symmetries. 

In recent literature, it has been shown that gauging subsystem symmetries of certain models can lead to fractonic phases of matter\cite{bravyi2011,yoshida2013,williamson2016,vijay2016,you2018b,vijay2016}. For the model considered here, gauging the subsystem symmetry in our non-local spin model leads to a (local) $\mathbb{Z}_2$ quantum double model, i.e., the toric code. In Appendix C, we explicitly gauge the subsystem symmetry of the effective projector Hamiltonian Eq. \ref{eq:2HOHam}, and show that this exactly leads to the toric code Hamiltonian. We can also see this in the continuum by setting $M^2= 0$ in Eq. \ref{eq:ContHiggs2}.  Because of this, the subsystem symmetry invariant model we constructed can be thought of as a $\mathbb{Z}_2$ quantum double model, where we have "un-gauged" the $\mathbb{Z}_2$ gauge symmetry along a certain direction, and reduced it to a $\mathbb{Z}_2$ subsystem symmetry.

It still remains to be seen how this construction generalizes to higher dimensions and different subsystem symmetries. In particular if there are general topological features of a $D$ dimensional system with a $d<D$ dimensional subsystem symmetry. The Mermin-Wagner theorem would seem to constrain $d<2$ for local theories, but exact details still need to be determined.  

Additionally, it is unknown what kind of classifications exist for these systems. It is known that topologically ordered systems can be classified based on their modular $S$ and $T$ matrices\cite{levin2005,chen2010,gu2015}. Since the systems discussed here are not invariant under modular transformations, we cannot define these $S$ and $T$ matrices in an analogous way. Currently, to our knowledge, there is no structure which performs to same role for the subsystem invariant systems, and so classification of these systems remains an open question.

\section*{Acknowledgements}
We would like to thank H. Goldman and M. Lin for useful conversations. TLH acknowledges support from the US National Science Foundation under grant DMR 1351895-CAR.

\bibliographystyle{apsrev4-1}

\begin{appendix}

\section{Ground State of Effective Projector Model of a Sphere}
\begin{figure}
\includegraphics[width=\linewidth]{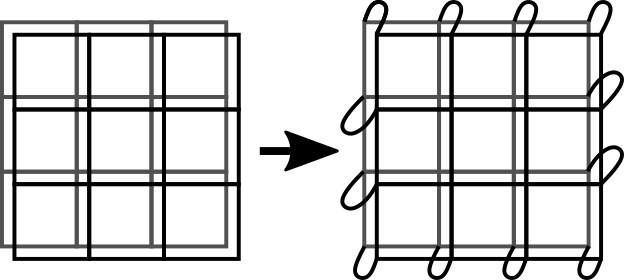}
\caption{ The process of sewing together two copies of a rectangular lattice at the edges to give it the topology of a sphere.}
\label{fig:LatticeSew}
\end{figure}

On a manifold with the topology of a sphere (genus 0), the effective projector Hamiltonian Eq. \ref{eq:2HOHam} has a unique ground state. We can show this explicitly by giving the lattice the topology of a sphere. To do this we will take two commuting copies of the system stacked on top of each other. The two copies are made into a sphere by ''sewing'' the copies together at the edges as shown in Fig. \ref{fig:LatticeSew}. After this, the Hamiltonian becomes
\beq
\nonumber H_s = &-& K \sum_{{\textbf{r}},\pm} \sigma^z_{{\textbf{r}},\pm} \sigma^z_{{\textbf{r}}+y,\pm} -K \sum_{{\textbf{r}} \parallel y-} \sigma^z_{{\textbf{r}},+} \sigma^z_{\textbf{r},-} \\\nonumber &-&K \sum_{{\textbf{r}} \parallel y+} \sigma^z_{{\textbf{r}},+} \sigma^z_{{\textbf{r}},-} - J \sum_{x_o}\prod_{{\textbf{r}} \in l(x_o)} \sigma^x_{{\textbf{r}},+} \sigma^x_{{\textbf{r}}+x,+}\sigma^z_{{\textbf{r}},-} \sigma^x_{{\textbf{r}}+x,-}
\\ &-&J \prod_{{\textbf{r}} \parallel x-} \sigma^x_{{\textbf{r}},+} \sigma^x_{{\textbf{r}},-} -J \prod_{{\textbf{r}} \parallel x+} \sigma^x_{{\textbf{r}},+} \sigma^x_{{\textbf{r}},-},
\label{eq:LatSphere}
\eeq
where the $\pm$ indexes the two stacked copies of the system, $y{-(+)}$ are the sites on the bottom (top) edge and $x{-(+)}$ are sites on the right (left) edge. As before, all terms present in the Hamiltonian commute. The ground state is thereby determined by minimizing each term individually.

Let us now count the constraints for this system. Consider a sphere created from sewing together two $L\times L$ lattices, leading to $2L^2$ spins. The first sum in Eq. \ref{eq:LatSphere} gives $2L^2-2L$ independent constraints. The second sum gives $L$ independent constraints. The third sum gives no independent constraints, since all terms in the third sum can be written as a product of terms in the first and second sums. The fourth sum gives $L-1$ independent constraints. The fifth term gives a single independent constraint. The final term gives no independent constraints, since it is a product of the terms in the fourth sum and the fifth term. So, in total we have $2L^2$ independent constraints. The ground state is thereby unique.

\section{Majorana Mean Field Theory}

Here will derive the Majorana mean field theory for the spin model Eq. \ref{eq:2HOHam} on a square $L\times L$ lattice. This will be done by decomposing each spin into 4 Majorana fermions, $\gamma^1$, $\gamma^2$, $\gamma^3$, and $\gamma^4$. These Majorana fermions obey the normal Majorana algebra, $\{\gamma^i,\gamma^j \} = 0$ and $\gamma^i\gamma^i = 1$. In terms of the Majorana fermions, the spin degrees of freedom can be rewritten as $\sigma^x = i\gamma^1\gamma^2$, $\sigma^y= i\gamma^1\gamma^3$, $\sigma^z = i\gamma^1\gamma^4$, with the local constraint that $\gamma^1\gamma^2\gamma^3\gamma^4 = 1$. It is straightforward to verify that the spin operators defined this way anti-commute. 

Let us now consider the terms in Eq. \ref{eq:2HOHam}. First there are the terms proportional to $K$, involving $\sigma^z_{\textbf{r}}\sigma^z_{\textbf{r}+\hat{y}}$. Due to the aforementioned constraint, $\sigma^z = i\gamma^1\gamma^4 = -i\gamma^2\gamma^3$. We can thereby write the $K$ terms of Eq. \ref{eq:2HOHam} as
\beq
-K\sigma^z_{\textbf{r}}\sigma^z_{\textbf{r}+\hat{y}} = -K\gamma^2_{\textbf{r}}\gamma^3_{\textbf{r}}\gamma^1_{\textbf{r}+\hat{y}}\gamma^4_{\textbf{r}+\hat{y}}.
\eeq Second, there are the terms proportional to $J$ $\prod_{r\in l(x_o)}\sigma^x_\textbf{r}\sigma^x_{\textbf{r}+\hat{x}}$. In terms of the Majorana fermions $\sigma^x = i\gamma^1\gamma^2 = -i\gamma^3\gamma^4$. We can thereby write the $J$ terms of Eq. \ref{eq:2HOHam} as
\beq
-J\prod_{r\in l(x_o)}\sigma^x_\textbf{r}\sigma^x_{\textbf{r}+\hat{x}} = -J\prod_{r\in l(x_o)}\gamma^3_\textbf{r}\gamma^4_\textbf{r}\gamma^1_{\textbf{r}+\hat{x}}\gamma^2_{\textbf{r}+\hat{x}}.
\eeq
where the product is over $l(x_o) = \{\textbf{r} = (x,y)| x=x_o, $ and $ 1\leq y\leq L\}$.

We will now introduce our mean field ansatz. First, note that $\gamma^3_\textbf{r}\gamma^4_{\textbf{r}+\hat{y}}$ and $\gamma^1_\textbf{r}\gamma^2_{\textbf{r}+\hat{y}}$ commute with all terms in the Hamiltonian. The $K$ terms can then be rewritten as
\beq
\nonumber -K\sigma^z_{\textbf{r}}\sigma^z_{\textbf{r}+\hat{y}} = -K_{1,2,\textbf{r}+\hat{y},\textbf{r}}\gamma^1_{\textbf{r}+\hat{y}}\gamma^2_{\textbf{r}} - K_{3,4,\textbf{r},\textbf{r}+\hat{y}} \gamma^3_{\textbf{r}}\gamma^4_{\textbf{r}+\hat{y}},\\
\label{eq:MF1}
\eeq
where $K_{1,2,\textbf{r}+\hat{y},\textbf{r}} \equiv K\langle \gamma^3_{\textbf{r}}\gamma^4_{\textbf{r}+\hat{y}}\rangle$ and $K_{3,4,\textbf{r},\textbf{r}+\hat{y}} \equiv K\langle \gamma^1_{\textbf{r}+\hat{y}}\gamma^2_{\textbf{r}}\rangle$. Eq. \ref{eq:MF1} is minimized by having  $\langle \gamma^3_{\textbf{r}}\gamma^4_{\textbf{r}+\hat{y}}\rangle = \langle \gamma^1_{\textbf{r}+\hat{y}}\gamma^2_{\textbf{r}}\rangle = 1$. This will correspond to a state with $\langle \sigma^z_{\textbf{r}}\sigma^z_{\textbf{r}+\hat{y}}\rangle = 1$ as expected from the spin-model analysis. 

More care must be taken with the $J$ terms due to the boundary conditions. If will be useful to rewrite the $J$ terms as 
\beq
\nonumber &-&J\prod_{r\in l(x_o)}\sigma^x_\textbf{r}\sigma^x_{\textbf{r}+\hat{x}} = -J\prod_{r\in l(x_o)}\gamma^3_\textbf{r}\gamma^4_\textbf{r}\gamma^1_{\textbf{r}+\hat{x}}\gamma^2_{\textbf{r}+\hat{x}}\\\nonumber = &-&J[\prod_{r\in l'(x_o)}\gamma^3_\textbf{r}\gamma^4_{\textbf{r}+\hat{y}}\gamma^2_{\textbf{r}+\hat{x}}\gamma^1_{\textbf{r}+\hat{x}+\hat{y}}]\\&\times& \gamma^4_{x_o,1}\gamma^1_{x_o+\hat{x},1}\gamma^3_{x_o,L}\gamma^2_{x_o+\hat{x},L},
\eeq
where the product is over $l'(x_o) = \{\textbf{r} = (x,y)| x=x_o $ and $ 1\leq y<L\}$. We can now rewrite $J$ terms in the mean field form
\beq
\nonumber &-&J\prod_{r\in l(x_o)}\sigma^x_\textbf{r}\sigma^x_{\textbf{r}+\hat{x}}\\\nonumber = &-&\sum_{\textbf{r}\in l'(x_o)}[J_{1,2,\textbf{r}+\hat{x}+\hat{y},\textbf{r}+\hat{x}}\gamma^1_{\textbf{r}+\hat{x}+\hat{y}}\gamma^2_{\textbf{r}+\hat{x}}+J_{3,4,\textbf{r},\textbf{r}+\hat{y}}\gamma^3_\textbf{r}\gamma^4_{\textbf{r}+\hat{y}} \\\nonumber &+& J_{4,1,\textbf{r},\textbf{r}+\hat{x}} \gamma^4_{x_o,1}\gamma^1_{x_o+\hat{x},1} + J_{3,2,\textbf{r},\textbf{r}+\hat{x}} \gamma^3_{x_o,L}\gamma^2_{x_o+\hat{x},L}.\\
\eeq 
In the above equation
\beq
\nonumber J_{1,2,\textbf{r}+\hat{x}+\hat{y},\textbf{r}+\hat{x}} &=& J\langle \gamma^1_{\textbf{r}+\hat{x}+\hat{y}}\gamma^2_{\textbf{r}+\hat{x}}\prod_{r'\in l(x_o)}\gamma^3_\textbf{r'}\gamma^4_\textbf{r'}\gamma^1_{\textbf{r'}+\hat{x}}\gamma^2_{\textbf{r'}+\hat{x}} \rangle\\\nonumber J_{3,4,\textbf{r},\textbf{r}+\hat{y}} &=& J\langle \gamma^3_\textbf{r}\gamma^4_{\textbf{r}+\hat{y}}\prod_{r'\in l(x_o)}\gamma^3_\textbf{r'}\gamma^4_\textbf{r'}\gamma^1_{\textbf{r'}+\hat{x}}\gamma^2_{\textbf{r'}+\hat{x}} \rangle\\\nonumber J_{4,1,\textbf{r},\textbf{r}+\hat{x}} &=& J\langle \gamma^4_{x_o,1}\gamma^1_{x_o+\hat{x},1}\prod_{r'\in l(x_o)}\gamma^3_\textbf{r'}\gamma^4_\textbf{r'}\gamma^1_{\textbf{r'}+\hat{x}}\gamma^2_{\textbf{r'}+\hat{x}}\rangle\\\nonumber J_{3,2,\textbf{r},\textbf{r}+\hat{x}} &=& J\langle \gamma^3_{x_o,L}\gamma^2_{x_o+\hat{x},L}\prod_{r'\in l(x_o)}\gamma^3_\textbf{r'}\gamma^4_\textbf{r'}\gamma^1_{\textbf{r'}+\hat{x}}\gamma^2_{\textbf{r'}+\hat{x}}\rangle .\\
\eeq

If we redefine $J_{1,2,\textbf{r}+\hat{y},\textbf{r}} \equiv J_{1,2,\textbf{r}+\hat{y},\textbf{r}}+ K_{1,2,\textbf{r}+\hat{y},\textbf{r}}$ and $J_{3,4,\textbf{r},\textbf{r}+\hat{y}} \equiv J_{3,4,\textbf{r},\textbf{r}+\hat{y}}+ K_{3,4,\textbf{r},\textbf{r}+\hat{y}}$, we can write the full mean field Hamiltonian as 
\beq
H = \sum_{\langle \textbf{r},\textbf{r}'\rangle }\sum_{i,j}J_{i,j,\textbf{r},\textbf{r'}}\gamma^i_{\textbf{r}}\gamma^j_{\textbf{r'}},
\eeq
where $J_{i,j,\textbf{r},\textbf{r'}}$ is non-zero for the following combinations:

$i=1$, $j=2$, $\textbf{r} = (x,y)$ with $0< y \leq L$ and $\textbf{r}' = \textbf{r}-\hat{y}$. 

$i=3$, $j=4$, $\textbf{r} = (x,y)$ with $0\leq y < L$ and $\textbf{r}' = \textbf{r}+\hat{y}$. 

$i=4$, $j=1$, $\textbf{r} = (x,0)$ with $0\leq x < L$ and $\textbf{r}' = \textbf{r}+\hat{x}$. 

$i=3$, $j=2$, $\textbf{r} = (x,L)$ with $0\leq x < L$ and $\textbf{r}' = \textbf{r}+\hat{x}$. 

Eq. \ref{eq:MajMF} gives the Majorana dimerization pattern shown in Fig. \ref{fig:4MajFig}. Each term in the mean field Hamiltonian Eq. \ref{eq:MajMF} commute. The ground state will then minimize each of the Majorana bi-linears in Eq. \ref{eq:MajMF}. Using the aforementioned identifications, it is clear that states which minimize the Majorana mean field Hamiltonian will also minimize each of the commuting elements in Eq. \ref{eq:2HOHam} as desired. 

\section{Gauging the Subsystem Symmetry}
\begin{figure}
\includegraphics[width=\linewidth]{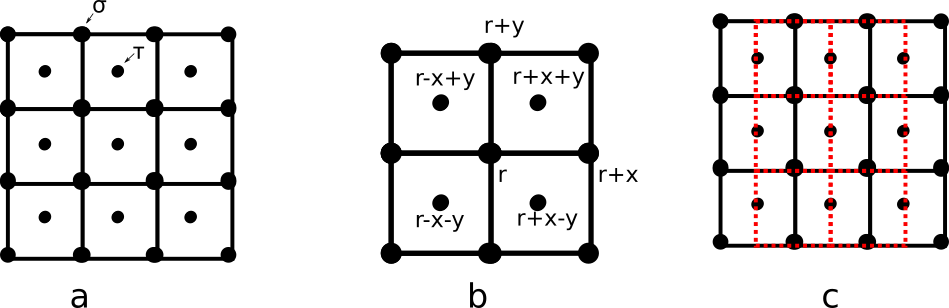}
\caption{ a) The additional spin 1/2 degrees of freedom $\tau$ added to gauge the subsystem symmetry. b) The new terms in the Hamiltonian after the addition of $\tau$. The labels correspond to Eq. \ref{eq:HamGauged}. c) The transformed lattice (red). on this lattice Eq. \ref{eq:HamGauged} is a $\mathbb{Z}_2$ lattice gauge theory.}
\label{fig:GaugeFig}
\end{figure}

Here, we will consider gauging the subsystem symmetry generated by $G[l(x_o)]$ in Eq. \ref{eq:SSGen1}. The paradigm of gauging a symmetry is to make the symmetry local. The local version of the symmetry generated by $G[l(x_o)]$ is generated by $\sigma^x_{\textbf{r}} \sigma^x_{{\textbf{r}}+x}$. This term does not commute with the Hamiltonian Eq. \ref{eq:2HOHam}. In order to make this local transformation a symmetry of the model, we will proceed in the standard fashion of adding in additional degrees of freedom, i.e., adding gauge fields. 

We will add additional spin-1/2 degrees of freedom $\tau$ at the center of each plaquette of the lattice as in Fig. \ref{fig:GaugeFig}a. After this we will change the first term of the Hamiltonian to 
\beq
-K\sum_{\textbf{r}}\sigma^z_{\textbf{r}} \sigma^z_{{\textbf{r}}+y}\tau^z_{\textbf{r} + x+y}\tau^z_{\textbf{r} - x-y},
\label{eq:HamGauged}
\eeq
where $\textbf{r} \pm x \pm y$ are the $\tau$ spins indicated in Fig. \ref{fig:GaugeFig}b. We can then flip $\sigma^z_{\textbf{r}}$ and $\sigma^z_{{\textbf{r}}+x}$ provided we also flip the two $\tau$ spins $\tau^z_{\textbf{r} + x+y}$ and $\tau^z_{\textbf{r} + x-y}$. So in Eq. \ref{eq:HamGauged} we have gauged the subsystem symmetry as desired. As the symmetry is now local, there is no need to include the non-local terms proportional to $J$. We can now include a term to energetically enforce invariance under the new local symmetry. This is done via the term
\beq
-J\sum_{\textbf{r}}\sigma^x_{\textbf{r}} \sigma^x_{{\textbf{r}}+x}\tau^x_{\textbf{r} + x+y}\tau^x_{\textbf{r} + x-y}.
\eeq

Let us now redefine the lattice as in Fig. \ref{fig:GaugeFig}c, and relabel $\tau \rightarrow \sigma$ (which should not cause confusion since the $\tau$ and $\sigma$ operators are defined on different lattice sites). After this relabeling we find that this Hamiltonian is 
\beq
H = -K \sum_{p} \prod_{l \parallel p} \sigma^x_l -J \sum_{v} \prod_{l \parallel v} \sigma^z_l, 
\eeq
where the sum is over the elementary plaquettes $p$ and vertices $v$. This is exactly the Hamiltonian for the deconfined $\mathbb{Z}_2$ lattice gauge theory, i.e.,  the toric code. Our spin model can thereby be identified as a $\mathbb{Z}_2$ lattice gauge theory where we have ''un-gauged'' the $\mathbb{Z}_2$ gauge symmetry into a subsystem symmetry.

\end{appendix}
\end{document}